\documentclass[reprint,amssymb,amsmath,aip,rsi]{revtex4-1}
\usepackage{docs}
\usepackage{bm}
\usepackage{graphicx}
\usepackage[colorlinks=true,linkcolor=blue]{hyperref}
\expandafter\ifx\csname package@font\endcsname\relax\else
 \expandafter\expandafter
 \expandafter\usepackage
 \expandafter\expandafter
 \expandafter{\csname package@font\endcsname}
\fi
\begin{document}
\title{Viscosity measurements in pulsed magnetic fields by using a quartz-crystal microbalance}
\author{T. Nomura}
\email{t.nomura@hzdr.de}
\affiliation{Hochfeld-Magnetlabor Dresden (HLD-EMFL), Helmholtz-Zentrum Dresden-Rossendorf, 01328 Dresden, Germany}
\affiliation{Institute for Solid State Physics, University of Tokyo, Kashiwa, Chiba 277-8581, Japan}
\author{S. Zherlitsyn}
\affiliation{Hochfeld-Magnetlabor Dresden (HLD-EMFL), Helmholtz-Zentrum Dresden-Rossendorf, 01328 Dresden, Germany}
\author{Y. Kohama}
\affiliation{Institute for Solid State Physics, University of Tokyo, Kashiwa, Chiba 277-8581, Japan}
\author{J. Wosnitza}
\affiliation{Hochfeld-Magnetlabor Dresden (HLD-EMFL), Helmholtz-Zentrum Dresden-Rossendorf, 01328 Dresden, Germany}
\affiliation{Institut f\"ur Festk\"orper- und Materialphysik, TU-Dresden, 01062 Dresden, Germany}
\date{\today}

\begin{abstract}
Viscosity measurements in combination with pulsed magnetic fields are developed by use of a quartz-crystal microbalance (QCM).
When the QCM is immersed in liquid, the resonant frequency, $f_0$, and the quality factor, $Q$, of the QCM change depending on $(\rho\eta)^{0.5}$, where $\rho$ is the mass density and $\eta$ the viscosity.
During the magnetic-field pulse, $f_0$ and $Q$ of the QCM are simultaneously measured by a ringdown technique.
The typical resolution of $(\rho\eta)^{0.5}$ is 0.5 \%.
As a benchmark, the viscosity of liquid oxygen is measured up to 55 T.
\end{abstract}
\maketitle

\section{Introduction}
A quartz-crystal microbalance (QCM) is an ultrasensitive device to detect changes of the mass \cite{Sauerbrey59,Schumacher90,FD97,Reviakine11}.
When an external mass is coupled to the surface of the quartz micro-membrane, the resonant frequency of the QCM, $f_0$, decreases.
Owing to the high quality factor, $Q$, of the QCM, a relative frequency change of 10$^{-9}$ can be resolved.
This offers various applications for monitoring purposes including surface deposition \cite{Kasemo80}, chemical reactions \cite{Alder83,Bruckenstein85}, and biomolecular adsorption \cite{Dixon08}.
Immersed in liquid, the QCM oscillates still in resonance, but with significantly decreased $f_0$ and $Q$.
The reduction of $f_0$ and $Q$ are proportional to $(\rho \eta)^{0.5}$, where $\rho$ is the mass density and $\eta$ is the shear viscosity of the liquid \cite{Bruckenstein85,Kanazawa85,Martin91,Rodahl95}.
Thus, the QCM is a useful device to investigate the viscous properties of liquids with high resolution.

In this paper, we present the viscosity-measurement technique in combination with pulsed magnetic fields up to 55 T.
Target materials include liquid oxygen \cite{Uyeda87,Uyeda88}, water \cite{Ghauri06,Chang06}, liquid crystals \cite{Yamagishi84,Ueda17}, and ferrofluids \cite{Odenbach04}, where the viscoelastic properties depend on the external magnetic field.
Potentially, this experimental technique can be applied for various measurements based on a QCM \cite{Kasemo80,FD97,Reviakine11,Schumacher90,Bruckenstein85,Alder83,Dixon08}.

The challenge of the high-field viscosity measurements is the short duration of the pulsed magnetic fields of about 10--100 ms.
It requires a sampling rate of the signal processing faster than 1 kHz.
$f_0$ and $Q$ of a QCM is typically monitored by use of a network analyzer.
However, the network analyzer cannot operate in such a short time scale.
In this research, we employ a ringdown dissipation monitoring technique \cite{Rodahl95,Dixon08,Reviakine11}, by which we can simultaneously monitor $f_0$ and $Q$ with a repetition rate of 10 kHz.
As a demonstration, we present the viscosity results for liquid oxygen up to 55 T.

\section{Experiment}
\subsection{Characterization of QCMs}
In this research, we use the commercial AT-cut quartz resonator chip NX3225SA (Nihon Dempa Kogyo Co. Ltd., fundamental resonance at 38.0 MHz) as a QCM.
The reproducibility of our results is confirmed with using another QCM, SEN-5P-H-10 (Tamadevice Co. Ltd., see supplementary material).
The advantages to use the commercial chip are the following.
First, the QCM is relatively stable and robust against mechanical vibrations and turbulences in liquid.
Second, it has high $f_0$ and $Q$, leading to higher sensitivity.
Here, high $f_0$ is favorable for a high repetition rate as well.
Third, the QCM is relatively small and cheap.
Inside the chip, a thickness-shear-mode quartz membrane with two electrodes is encapsulated.
Figure 1(a) shows a photograph of the chip after removing the hermetic seal.
The quartz membrane is fixed only at the edge and the active areas of the QCM are free-standing from the substrate.
When the chip is immersed in liquid, both sides of the membrane have contact with the liquid environment.

The resonance properties of the QCM were characterized by a vector network analyzer mini-VNA Tiny (WiMo Antennen \& Elektronik GmbH).
Return loss spectra of the QCM in vacuum at 93 K and in liquid oxygen at 77.4 K are shown in Fig. 1(c).
In vacuum, a sharp resonance peak is observed at 114.2 MHz, which corresponds to the third harmonic of the oscillator.
In liquid oxygen, the resonance broadens and shifts to lower frequency due to the mass-loading effect and viscosity contribution.
In vacuum, weak satellite peaks are seen at a frequency about 0.1 MHz higher than the main peak.
In the liquid, they are not resolvable anymore.
The irrelevance of satellite peaks is important to obtain reproducible results without interference.

\begin{figure}[tb]
\centering
\includegraphics[width=0.95\linewidth]{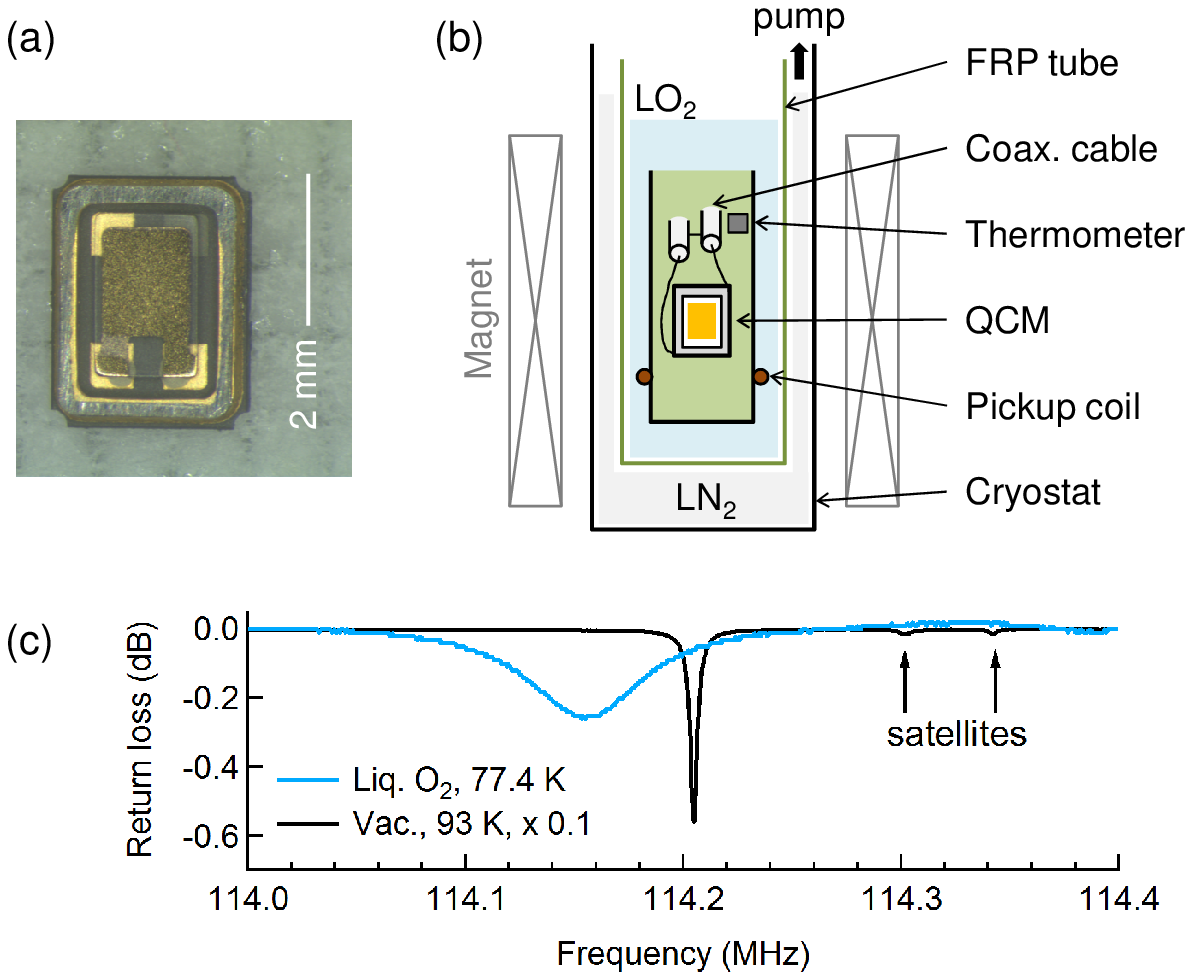}
\caption{
(a) Photograph of the used QCM.
(b) Schematic experimental setup with the QCM immersed in liquid oxygen.
(c) Return loss spectra of the QCM at frequencies around the third harmonic.
The black and cyan curves represent the results in vacuum (93 K) and in liquid oxygen (77.4 K), respectively.
The black curve is multiplied by 0.1.
The curves are shifted to be zero at 114.0 MHz.
} 
\end{figure}

\subsection{QCM with dissipation monitoring}
Here, we introduce the measurement setup based on the ringdown technique, the so-called QCM with dissipation monitoring (QCM-D) \cite{Rodahl95,Dixon08,Reviakine11}.
In this technique, resonant oscillatory motion of the QCM is intermittently excited by an RF generator.
Even after the RF generator is switched off, the mechanical oscillation of the QCM persists with exponential decay.
This residual oscillation directly reflects the resonant properties of the QCM, characterized by $f_0$ and $Q$.

The schematic measurement diagram is shown in Fig.~2(a).
In this study, we choose the transmission geometry \cite{Spencer66} rather than the reflection geometry to avoid noise originating from the switching device or directional coupler.
Each side of the quartz membrane is connected to the core of a coaxial cable.
One coaxial cable is used to excite the resonant oscillation of the QCM and another is used to monitor the free-decay behavior.
An RF generator with pulse modulation (SG382, Stanford Research Systems) is used to excite the QCM.
The excitation frequency is tuned to the resonant frequency of the QCM within a maximum deviation of 0.1 \%.
Here, higher harmonics of the resonant mode can be selectively excited by tuning the frequency.
A slight deviation of the excitation frequency from $f_0$ does not affect the measurements, although the amplitude of the oscillation is reduced.
The typical RF pulse duration and the pulse repetition rate are 4 $\mu$s and 10 kHz, respectively.
The excitation voltage is $V_\mathrm{pp}=0.5$ V.
The transmitted signal is recorded by an oscilloscope (HDO6104A, Teledyne LeCroy) with a sampling rate of 1.25 GS/s.
The 10 MHz timebase of the RF generator and the oscilloscope is synchronized to fix the phase of the RF pulse for each excitation.

\begin{figure}[tb]
\centering
\includegraphics[width=0.99\linewidth]{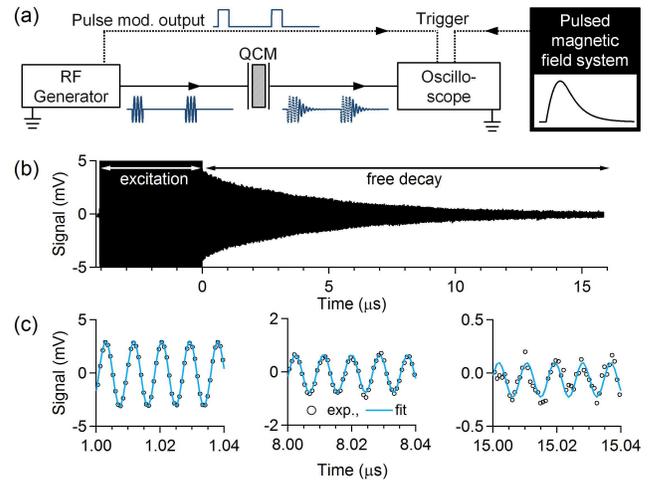}
\caption{
(a) Schematic measurement diagram of the QCM with dissipation monitoring.
Measurement and trigger lines are shown by solid and dotted lines, respectively.
(b) Typical decay curve of the QCM in liquid oxygen.
(c) Enlarged free-decay signal with the fit based on Eq. (1). 
Results for liquid oxygen at 77 K ($f_0=114.153$ MHz, $Q=1703$) are shown.
} 
\end{figure}

A typical decay curve is shown in Fig. 2(b).
The experiment was done for liquid oxygen at 77 K with an excitation frequency of 114.1 MHz (third harmonic).
The RF generator excites the QCM from $-4$ to 0 $\mu$s using pulse modulation.
During the excitation, the recorded voltage is intervened by transmission and crosstalk noise ($\sim$15 mV).
Only the freely decaying oscillation after the excitation (time, $t>0$) is used for the analysis.
The freely decaying voltage as a function of time, $A(t)$, can be described by an exponentially damped sinusoid \cite{Rodahl95,Dixon08,Reviakine11},
\begin{equation}
A(t)=A_0 e^{-t/\tau} \mathrm{sin}(2\pi f_0t+\phi) + \mathrm{constant},
\end{equation}
where $A_0$ is the initial amplitude, $\tau$ is the decay-time constant, and $\phi$ is the phase.
$Q$ is expressed by using $\tau$ as,
\begin{equation}
Q=\pi f_0 \tau.
\end{equation}
Figure 2(c) shows that Eq.~(1) fits the data very well.
If the satellite resonance peaks are located near $f_0$, the fit does not work due to beating.
The obtained $f_0$ and $Q$ do not depend on the excitation voltage, as checked between 0.1 and 2 V.
This indicates that the QCM is working in a linear regime without any spurious caused by cavitations or surface slips.

\subsection{Sample preparation and pulsed-magnetic-field experiment}
Figure 1(b) shows the schematic experimental setting. 
A fiber-reinforced plastic (FRP) tube is filled with oxygen gas (99.999 \%) and cooled by liquid nitrogen.
The temperature, $T$, is controlled by pumping the liquid nitrogen and monitored by a RuO$_2$ thermometer.
The pressure in the sample space is always kept at the vapor pressure of the liquid oxygen.
The transient magnetic field, $H$, is measured by using a pickup coil located near the QCM.

The experiments were performed in the Hochfeld-Magnetlabor Dresden (HLD-EMFL).
The typical maximum field and whole pulse duration were 60 T and 150 ms, respectively.
The same experimental procedure can, in principle, be applied for other multiple-coil nondestructive pulsed magnets to reach even higher magnetic fields \cite{Zherlitsyn13}.

The QCM-D data were recorded by the oscilloscope in a segmental sampling mode, where the sampling event occurred for each excitation.
Typically, one sampling segment records 20 $\mu$s and repeats after an 80 $\mu$s break.
In other words, the measurement is repeated at 10 kHz with a duty-cycle ratio of 20 \%.
The sampling is triggered by the pulsed-magnetic-field system and repeated by the TTL signal from the pulse modulation of the RF generator [Fig. 2(a)].
The segmental sampling mode reduces data size and the computational complexity of the analysis.

\section{Result and discussion}
\subsection{Temperature dependence}
First, the performance of the QCM-D was tested by measuring the temperature dependence of $f_0$ and $Q$ in liquid oxygen at zero field [Figs. 3(a)-(b)].
With decreasing temperature, $f_0$ and $Q$ monotonically decrease.
This is due to the increased density, $\rho$, and viscosity, $\eta$, which enhance the mass-loading effect.
Note, that the changes of the density and elastic constant of quartz are negligible in this temperature range.

The changes of $f_0$ and $Q$ are proportional to the coupled mass on the QCM, which is proportional to $(\rho \eta)^{0.5}$ \cite{Bruckenstein85,Kanazawa85}.
By using reported data for $\rho(T)$ \cite{Pentermann78} and $\eta(T)$ \cite{Hellemans70}, we are able to plot $f_0$ and $Q$ as a function of $(\rho \eta)^{0.5}$ [Figs. 3(c)-(d)].
$f_0$ and $Q$ scale almost linearly with $(\rho \eta)^{0.5}$ in agreement with earlier studies \cite{Bruckenstein85,Kanazawa85,Martin91,Rodahl95}.
Slight deviation from linearity might be due to the relatively large frequency shift $\Delta f/f_0 \sim 0.1$ \%.
The quadratic fit gives better results, which is used for the conversion of the pulsed-field data, from $f_0$ and $Q$ to $(\rho \eta)^{0.5}$.

\begin{figure}[tb]
\centering
\includegraphics[width=0.99\linewidth]{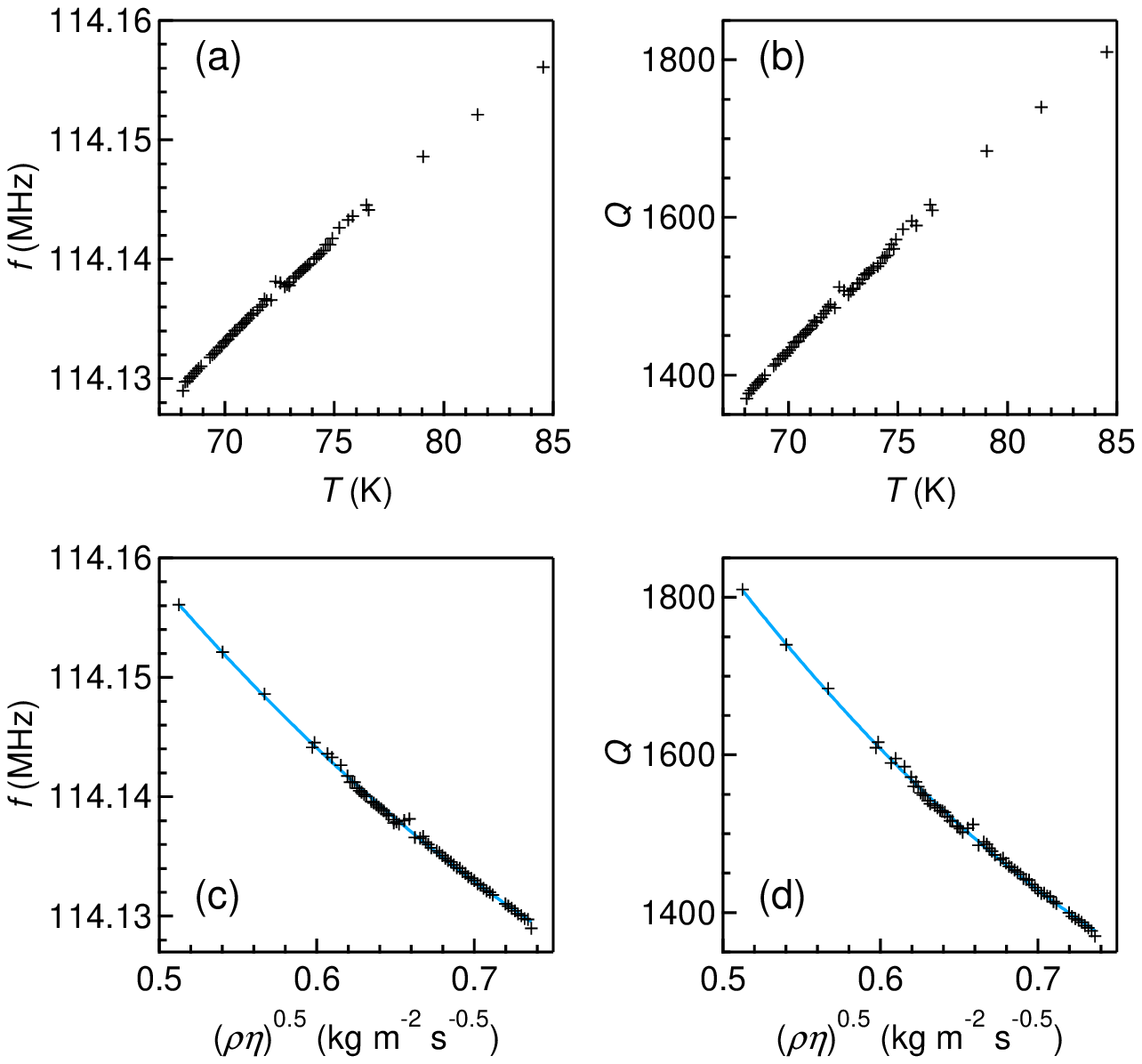}
\caption{
Temperature dependences of (a) $f_0$ and (b) $Q$ in liquid oxygen.
$(\rho \eta)^{0.5}$ dependences of (c) $f_0$ and (d) $Q$.
The cyan lines show quadratic fits.
} 
\end{figure}

\subsection{Pulsed-field results}
Figures 4(a) and 4(b) show the $H$ dependence of $f_0$ and $Q$ at selected temperatures.
The noise level for $f_0$ and $Q$ is 1 ppm and 1~\%, respectively.
Figure~4(c) shows the extracted $(\rho \eta)^{0.5}$, which is independently obtained from $f_0$ and $Q$ by using the fits shown in Figs. 3(c)-(d) for the conversion.
The results obtained from $f_0$ and $Q$ agree well.
This suggests that there is no interference from the satellite resonance peaks [Fig. 1(c)], and the $(\rho \eta)^{0.5}$ dependences of $f_0$ and $Q$ hold also in magnetic fields.
The noise level of $(\rho \eta)^{0.5}$ obtained from $f_0$ and $Q$ is 0.5~\% and 1~\%, respectively.

With increasing magnetic field, $(\rho \eta)^{0.5}$ monotonically decreases.
The decrease of $(\rho \eta)^{0.5}$ at 50 T is 2.9(3)~\% for all selected temperatures between 68.1 and 84.5 K [Fig. 4(c)].
Here, the decrease of $(\rho \eta)^{0.5}$ cannot be explained by a $T$ change during the adiabatic magnetization (magnetocaloric effect, MCE).
The adiabatic temperature change of liquid oxygen is approximately 0.5 K at 50 T \cite{Nomura17}.
However, if the MCE would be relevant, $T$ and $(\rho \eta)^{0.5}$ have to show hysteresis at quasi-isothermal conditions \cite{Kihara13}.
Therefore, the decrease of $(\rho \eta)^{0.5}$ is attributed to the magnetic-field dependences of $\rho$ and $\eta$.

To separate the contributions from $\rho$ and $\eta$, two different QCMs with controlled surface roughness are necessary \cite{Martin94}.
Here, as a rough estimate, we simply take the reported $\rho(H)$ and extract $\eta(H)$.
The density change of liquid oxygen was reported as $\Delta \rho/\rho(H) = -3.5\times10^{-6} H^2$ (\%), where $H$ is in Tesla \cite{Uyeda87}.
This value was obtained for liquid oxygen at 77 K in magnetic fields up to 8 T.
The decrease of $\rho$ is due to the antiferromagnetic (AFM) exchange striction \cite{Uyeda87,Bussery93}. %, $J \approx -30$ K at 3.2 ang
The external magnetic field works as a repulsive force for the O$_2$ molecules, since weaker AFM interaction is favored for polarized magnetic moments.
By extrapolating the quadratic dependence, we estimate that $\Delta \rho/\rho(\mathrm{50\ T}) = -0.9$~\%.
To explain the measured change of $(\rho \eta)^{0.5}$, $\Delta \eta/\eta(\mathrm{50\ T}) = -4.9(6)$~\% has to be assumed.

\begin{figure*}[tb]
\centering
\includegraphics[width=0.99\linewidth]{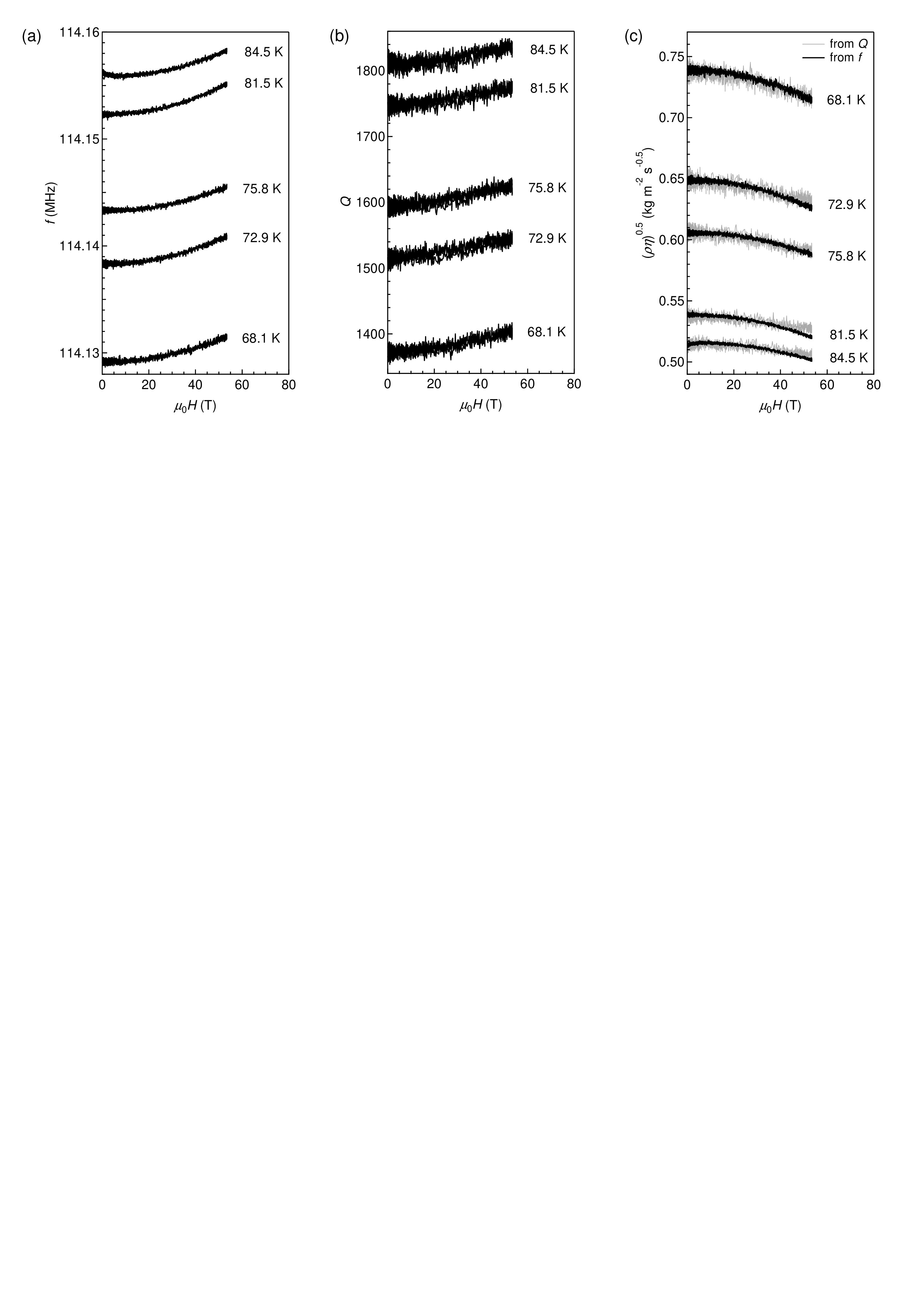}
\caption{
Magnetic-field dependences of (a) $f_0$ and (b) $Q$ in liquid oxygen at selected temperatures.
(c) $(\rho \eta)^{0.5}$ is independently obtained from $f_0(H)$ (black) and $Q(H)$ (gray).
} 
\end{figure*}

In the free volume theory, the viscosity depends on the free volume of the molecules \cite{Hellemans70,Herrman76}.
Therefore, $\eta$ can be scaled by $\rho$ in a limited temperature range.
In Fig.~5, we plot the viscosity data obtained in high-pressure, $P$, experiments at isothermal conditions \cite{Itterbeek66,Haynes77,Grevendonk68}.
Here, we used $\rho(P)$ data \cite{Itterbeek60} to convert $P$ to $\rho$.
The gray dotted line shows the viscosity at saturated vapor pressures \cite{Herrman76}.
In this temperature range from 70 to 90 K, $\eta$ scales well with $\rho$.

Our results obtained at 75.8 and 81.5 K are plotted as black lines.
Here, we extracted $\eta(H)$ by using the experimentally obtained $(\rho \eta)^{0.5}$ [Fig. 4(c)] and $\rho(H)$ which is estimated by extrapolating the data at 77 K from literature \cite{Uyeda87,Pentermann78}. 
$\eta(P)$, $\eta(T)$, and $\eta(H)$ scale well with $\rho$.
Therefore, the observed change of $(\rho \eta)^{0.5}$ and the obtained $\eta(H)$ are reliable on a quantitative level.

\begin{figure}[tb]
\centering
\includegraphics[width=0.99\linewidth]{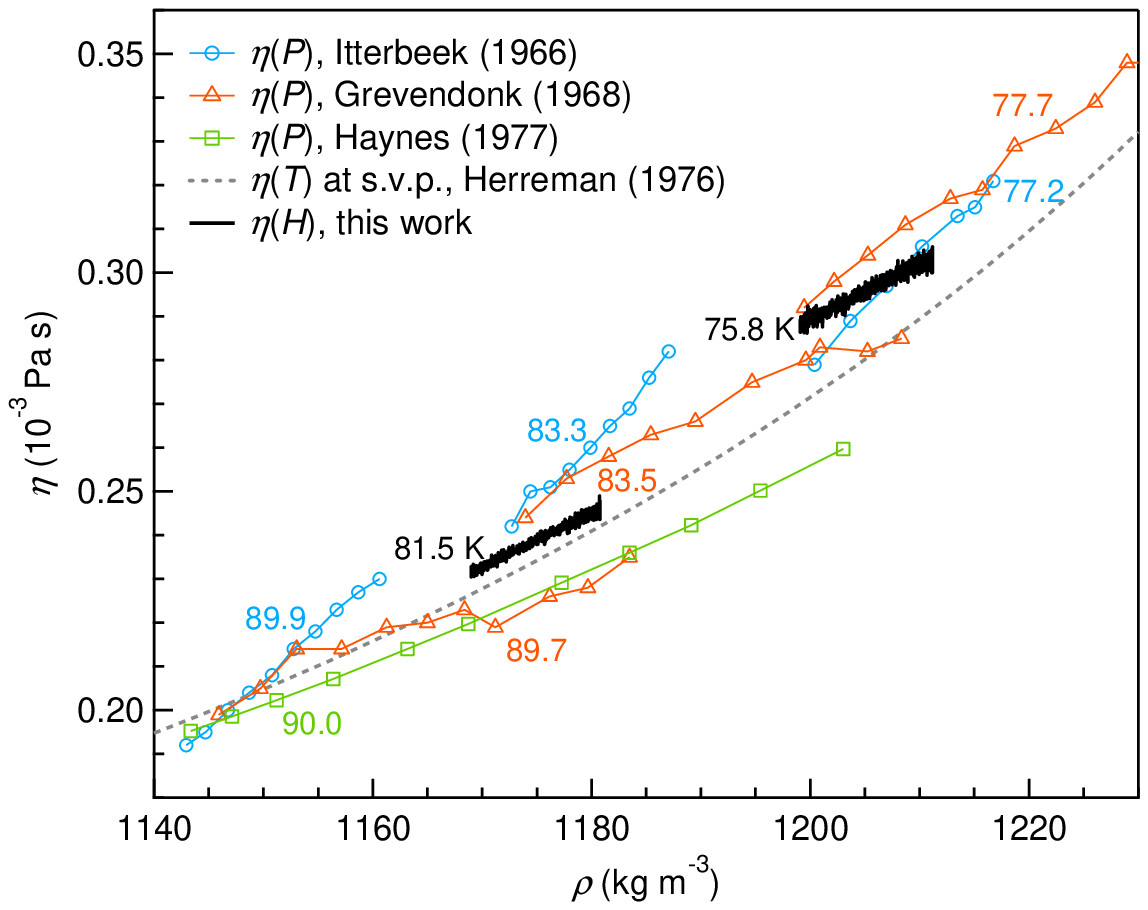}
\caption{
Viscosity of liquid oxygen scaled by density.
The symbols show the data obtained from $\eta(P)$ at isothermal conditions \cite{Itterbeek66,Haynes77,Grevendonk68}. 
$\eta(T)$ at saturated vapor pressures (s.v.p.) is shown by the dotted line \cite{Herrman76}.
The black lines show the results obtained from $\eta(H)$ in this work.
Temperatures are denoted for each $P$ or $H$ sweep.
See text for details.
} 
\end{figure}

\section{Conclusion}
We have developed a viscosity-measurement technique in pulsed magnetic fields.
A QCM with ringdown technique enables sensitive and fast measurements of $(\rho \eta)^{0.5}$.
This experimental procedure can be applied for other researches using QCMs with pulsed magnetic fields.

\section*{Supplementary material}
The supplementary material shows the reproducibility of the results using another model of a QCM.

\section*{Acknowledgments}
We appreciate fruitful discussions with Y. H. Matsuda and T. C. Kobayashi.
This work was partly supported by the BMBF via DAAD (project-id 5745-7940), and by HLD at HZDR, member of the European Magnetic Field Laboratory. 
T.N. was supported by the JSPS through a Grant-in-Aid for JSPS Fellows.

\end{document}